\documentclass[largeformat]{interact}
\usepackage{amsmath}
\usepackage{braket}
\usepackage[utf8]{inputenc}
\usepackage{graphicx}
\usepackage{color}
\usepackage{multirow}
\usepackage{comment}
\usepackage{subcaption}
\usepackage[Euler]{upgreek}
\usepackage{makecell}
\usepackage{bm}
\usepackage{hyperref}
\usepackage{physics} %
\usepackage{mathrsfs} %
\usepackage{mycommands}

\usepackage[numbers,sort&compress,merge]{natbib}%
\bibpunct[, ]{(}{)}{,}{n}{,}{,}%
\renewcommand\bibfont{\fontsize{10}{12}\selectfont}%
\renewcommand\citenumfont[1]{\textit{#1}}%
\renewcommand\bibnumfmt[1]{(#1)}%

\begin{document}
\articletype{Festschrift in Honour of Fr\'ed\'eric Merkt}
\title{Multidimensional tunnelling of molecules aligned by strong electric fields}

\author{
\name{J. Amira Geuther, Marit R. Fiechter, Jeremy O. Richardson\textsuperscript{a}\thanks{CONTACT J.O. Richardson. Email: jeremy.richardson@phys.chem.ethz.ch}}
\affil{Institute of Molecular Physical Science, ETH Z\"urich, 8093 Z\"urich, Switzerland}
}
\maketitle

\begin{abstract} 
Strong electric fields can be used to align molecules. %
However, a non-polar molecule such as \ce{H2} has no preference for %
its orientation.
There are thus two equivalent configurations with equal energy separated by a potential-energy barrier.
Quantum mechanically, the molecule can tunnel between these configurations resulting in a tunnelling splitting, which in the case of \ce{H2}, is the same as the ortho--para splitting. In this work, we generalize semiclassical instanton theory to calculate the energy splitting of molecules in electric fields in full dimensionality. This goes beyond a perturbative treatment of the field and takes into account changes in molecular geometry during the tunnelling process which influence its electrical properties and can have a significant impact on the result. We first study the case of H$_2$ in a static electric field and then show how it can be applied to larger polar molecules subjected to oscillating electric fields, where we find that even large-amplitude heavy-atom tunnelling can lead to observable splittings.

\end{abstract}


\section{Introduction}
\begin{figure}
    \centering
    \includegraphics[width=0.5\linewidth]{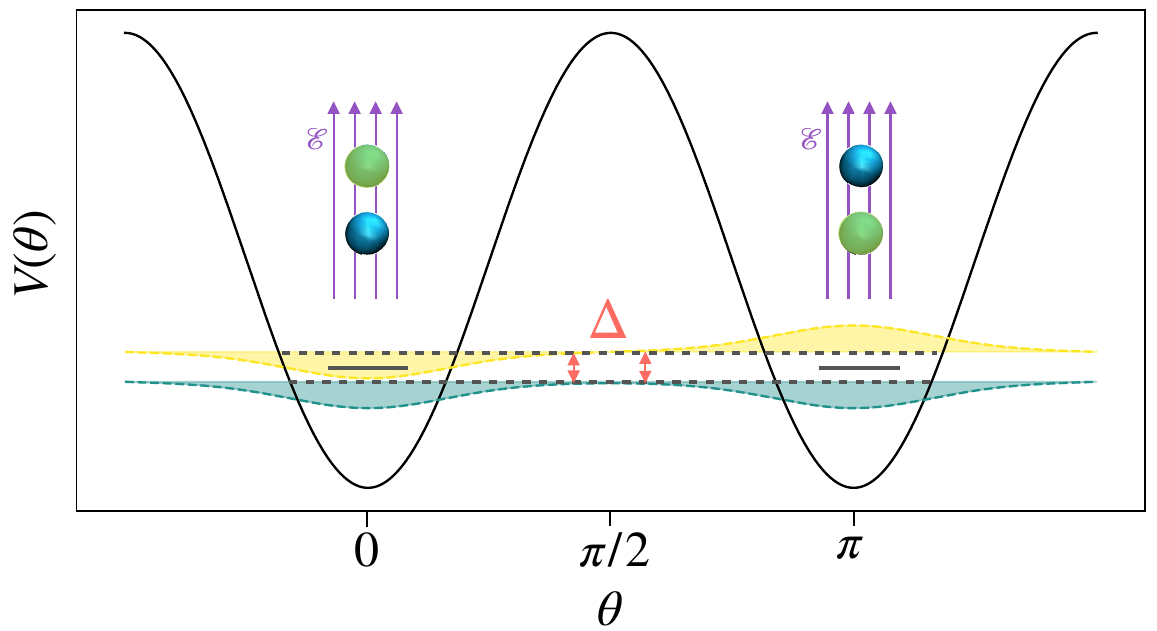}
    \caption{Illustration of the tunnelling splitting for a hydrogen molecule aligned in a strong electric field. The indistinguishable atoms are coloured to clearly illustrate the two degenerate states.
    Quantum mechanically, the molecule can tunnel between these, resulting in two delocalized eigenstates marked by the dashed lines. The quantity we seek is the splitting of these two levels, $\Delta$.
    Note that the angle is only defined on the interval $\theta=[0,\pi]$, although for illustration purposes we show mirror images of the potential and wavefunctions outside this range.}
    \label{fig:ts}
\end{figure}

Over the past few decades, experimental and theoretical interest in aligning molecules using strong electric fields has sparked much discussion on new observable phenomena \cite{friedrich_polarization_1995,friedrich_alignment_1995,friedrich_electro-optical_2022,nakagami_optimal_2008,lemeshko_manipulation_2013,chatterley2017three}.
Theoretical studies have suggested that both polar and symmetric polarizable molecules aligned in strong fields can exhibit pendular motion, as well as an energy-level splitting due to tunnelling \cite{friedrich_polarization_1995}.
To understand how tunnelling splittings can appear in molecules in strong electric fields, it is instructive to look at the 
case of a hydrogen molecule aligned in a static homogeneous electric field. While H$_2$ lacks a permanent dipole moment, it is polarizable, such that it prefers to align with the field. As a result, the system has two minimal potential energy configurations, corresponding to the two equivalent orientations.\footnote{Note the distinction between alignment and orientation: alignment describes the axis of a molecule being parallel to a field, regardless of direction, while orientation means that the molecule points in a specific direction along the field.}
These two states are degenerate and are coupled via quantum-mechanical tunnelling, resulting in a energy-level splitting as depicted in Fig.~\ref{fig:ts}.

The two eigenstates can be identified with the ortho and para spin isomers of the hydrogen molecule: in para-H$_2$, the wavefunction must be spatially symmetric, making it the lower-energy state, while in ortho-H$_2$, the wavefunction is spatially antisymmetric and thus higher in energy. The tunnelling splitting therefore corresponds to the ortho--para energy difference. Although this splitting is challenging to observe directly (due to the conservation of nuclear spin symmetry in most spectroscopic experiments), %
methods have been developed to measure forbidden transitions indirectly through transitions to intermediate states
\cite{beyer_determination_2019}.

To align molecules in an electric field, one can in principle use homogeneous static fields \cite{li_brute_1998}. However, much larger field strengths can be reached by lasers \cite{fleischer_molecular_2012}, where the electric field oscillates in time. %
Typically, this oscillation is so fast that in the multipole expansion all terms containing odd powers of the electric field (\emph{e.g.} the dipole interaction) average out to zero, such that the effective field only interacts with the polarizability (and other higher-order terms containing even powers of the field) \cite{friedrich1999manipulating}. An interesting consequence of this is that in fast oscillating fields, even polar molecules experience a symmetric double-well potential and will therefore exhibit a tunnelling splitting (whereas in a static field, polar molecules would have a preferred orientation, in which their dipole moment is pointing along the direction of the field).

There exists a range of theoretical methods for calculating tunnelling splittings in molecular systems, including fully quantum-mechanical approaches, where the Schrödinger equation is discretized on a spatial grid and solved for its eigenstates \cite{Hammer2011malonaldehyde,Schroeder2011malonaldehyde,Carrington2017quantum,Lauvergnat2023malonaldehyde,Simko2025trimer}.
However, this becomes impractical for high-dimensional problems without artificially reducing the dimensionality \cite{Kamarchik2009reduced}. %
This highlights the need for alternative methods based on reliable approximations.

One such alternative is instanton theory, which relies on a semiclassical approximation within the path-integral formalism of quantum mechanics \cite{Miller1971density,Uses_of_Instantons,Benderskii,Kleinert,ankerhold_quantum_2007}. This method involves locating the optimal tunnelling pathway, which can be implemented with the ring-polymer formalism to provide a practical framework for calculating tunnelling splittings in molecular systems
\cite{Perspective,richardson_ring-polymer_2018,marquardt_chapter_2021,tunnel,octamer,hexamerprism,formic,i-wat2,chiral,asymtunnel,tropolone,TransferLearning,Videla2023tunnel,AnharmInst,fiechter_ring-polymer_2025,Cvitas2016instanton,Cvitas2018instanton,pentamer,Erakovic2020instanton,Erakovic2021trimer,Tokic2025hexamer}.

In this work, we extend instanton theory to treat molecules in electric fields. This comes with an added complexity,
due to the presence of rotational invariance of the tunnelling pathway, which gives rise to an extra zero frequency in addition to the well-known permutational mode.
In the following, we introduce the theoretical framework needed to evaluate the tunnelling splitting using instanton theory and show how to address this new issue. We then study H$_2$ in an electric field.
First, we introduce a model system constructed using a multipole expansion to validate the derived theory.
After this, we employ a non-perturbative treatment of the field by including it explicitly in \textit{ab initio} electronic-structure calculations evaluated ``on the fly'' as we optimize the instanton. %
Finally, we showcase the method by studying the reorientational tunnelling in full dimensionality of a polyatomic polar molecule, formaldehyde, in an intense high-frequency laser field.
In this way, we demonstrate that the tunnelling of heavy atoms may give rise to observable tunnelling splittings, which will hopefully inspire experiments to measure this intriguing phenomenon.

\section{Theory}

A molecule in an electric field can be described by the full molecular Hamiltonian with an added scalar potential. In atomic units, this is given by
\begin{align}\label{eq:Ham}
    \hat{H}_\mathrm{mol} = &-\sum_{i}\frac{\nabla_{i}^2}{2} -\sum_{a}\frac{\nabla_a^2}{2m_{a}}  - \sum_a\sum_i
 \frac{Z_a}{\left| \mathbf{R}_a - \mathbf{r}_i \right| } +\sum_{i}\sum_{j > i}
 \frac{1}{ \left| \mathbf{r}_i - \mathbf{r}_j \right| } \nonumber\\&+\sum_{a}\sum_{b > a}
 \frac{Z_a Z_b}{ \left| \mathbf{R}_a - \mathbf{R}_b \right| }+ \sum_{a} Z_a\,\phi(\mathbf{R}_a) -\sum_{i} \,\phi(\mathbf{r}_i) ,
\end{align}
where %
indices $a$ and $b$ %
refer
to nuclei and $i$ and $j$ to electrons. %
The Hamiltonian contains the standard kinetic energy operators, where %
$m_a$ is the mass of nucleus $a$. Moreover, there are electron--electron, electron--nucleus and nucleus--nucleus interactions with nuclear and electronic positions $\mathbf{R}_a$ and $\mathbf{r}_i$ and nuclear charges $Z_a$. Lastly, the electric potential %
$\phi(\mathbf{R}_a) = -\mathbf{E}\cdot\mathbf{R}_a$ is
defined in terms of the static electric field $\mathbf{E}=(0,0,\mathscr{E})$.

We will employ the Born--Oppenheimer approximation to separate the electronic and nuclear degrees of freedom.
Using the illustrative case of a hydrogen molecule in an electric field, we can write the resulting potential energy surface (PES) in spherical polar coordinates $(r,\theta,\varphi)$, where $r$ is the internuclear distance, %
 $\theta$ is the angle between the molecular axis and the electric field, and $\varphi$ is the azimuthal angle.
This gives 
\begin{equation}
        \hat{H} = -\sum_a\frac{\nabla_a^2}{2m_a}  + V(r,\theta) .
\end{equation}
Note that, due to the symmetry of the problem, there is no dependence on the $\varphi$ coordinate.
However, along the $\theta$ degree of freedom, the potential energy surface has two minima corresponding to $\theta = 0$ and $\theta =\pi$ and a maximum at $\theta=\pi/2$, as shown schematically in Fig.~\ref{fig:ts}. We wish to study the dynamics of the molecule as it tunnels between the two degenerate states through the barrier.

\subsection{Tunnelling splittings in the path-integral framework}
In principle, the ground-state tunnelling splitting is defined as the energy difference between the lowest two eigenstates obtained by solving the Schrödinger equation. %
However, as mentioned before, this is prohibitively expensive for most molecular systems. We therefore rephrase the problem in terms of path integrals. As we will see later, in this framework it is straightforward to take a semiclassical approximation, which dramatically simplifies the problem. This allows us to extract reasonably accurate values for the tunnelling splitting from full-dimensional calculations for an acceptable computational cost \cite{Perspective}.

To define the tunnelling splitting in a path-integral framework, we first express it in terms of the imaginary-time propagators between the minima of the two wells \cite{Uses_of_Instantons,AnharmInst}. %
In particular, we %
take the low-temperature ($\beta\rightarrow\infty$) limit such that only the two lowest-energy eigenstates contribute to the propagator.
The tunnelling splitting, $\Delta$, can then be obtained in terms of the ratio of the propagators in imaginary time $\beta\hbar$, one which connects the two minima and the other which ends where it started: %
\begin{equation} \label{eq:Deltafromprop}
    \Delta = \lim_{\beta \rightarrow \infty}  \frac{2}{\beta} \, \arctanh \left( \frac{\bra{q_{\mathrm{fin}}}\mathrm{e}^{-\beta \hat{H}}\ket{q_{\mathrm{init}}}}{\bra{q_{\mathrm{init}}}\mathrm{e}^{-\beta \hat{H}}\ket{q_{\mathrm{init}}}} \right),
\end{equation}
where %
$ q=\{\sqrt\frac{m_a}{m} \mathbf{R}_a\}$ is an $f$-dimensional vector ($f=3\times\text{number of atoms}$) containing the mass-weighted coordinates of all the atoms in the molecule (with a reference mass $m$, \emph{e.g.} equal to the mass of a hydrogen atom) and $q_{\mathrm{init}}$ and $q_{\mathrm{fin}}$ refer to the configurations of the two minima (\textit{i.e.} one with $\theta=0$ and the other with $\theta=\pi$).

Thus, in order to calculate the tunnelling splitting, we need to evaluate the propagators. This can be carried out efficiently using ring-polymer instanton theory, as explained in the following. For this, we first rewrite the propagators in terms of discretized path integrals and then evaluate them with a semiclassical approximation.

\subsection{Semiclassical approximation to path integrals} \label{sec:semiclassical}
The propagator can be expressed in terms of a path integral, discretized into $N$ imaginary-time intervals \cite{Feynman}:
\begin{equation}
    \bra{q'}\eu{-\beta \hat{H}}\ket{q''}= \lim_{N \rightarrow \infty} A_N \int \eu{-S_N(\bm{q})/\hbar} \, \dd\bm{q} ,\label{eq:propS}
\end{equation}
where the action of the path is given by
\begin{equation} \label{eq:action}
    S_N(\bm{q}) = \sum_{n=1}^{N} \frac{m}{2 \tau_N} \lVert q_n - q_{n-1} \rVert^2 
    + \tau_N \left[ \thalf V(q_0) + \sum_{n=1}^{N-1} V(q_n) + \thalf V(q_N) \right].
\end{equation}
Here, the end beads are fixed to the geometries $q_0\equiv q'$ and $q_N\equiv q''$ (\emph{e.g.} $q_\mathrm{init}$ or $q_\mathrm{fin}$ as appropriate), and
$\bm{q} = (q_{1},\dots,q_{N-1})$ are the positions of the $N-1$ free beads.
Additionally, $A_N$ is the path-integral normalization factor
\begin{equation}
    A_N = \left( \frac{m}{2\pi \hbar \tau_N }\right)^{Nf/2},
\end{equation} with $\tau_N = {\beta \hbar}/{N}$.
Identifying that $U_N(\bm{q})=S_N(\bm{q})/\tau_N$ is equivalent to the potential of a fixed-ended polymer chain with $N$ springs, we note that the quantum-mechanical propagator [Eq.~\eqref{eq:propS}] is isomorphic to a classical statistical mechanical configuration integral over a fixed-ended linear polymer in thermal equilibrium. %

In order to evaluate the integrals in Eq.~\eqref{eq:propS}, we expand the action to second order around its minimum $\Tilde{\bm{q}}$ to give $(N-1)f$ Gaussian integrals that can easily be evaluated analytically \cite{BenderBook}:
\begin{align}
    \int \eu{-S_N(\bm{q})/\hbar} \, \dd \bm{q} &\simeq \int \eu{-S_N(\Tilde{\bm{q}})/\hbar - (\bm{q} -\Tilde{\bm{q}})^T \cdot \bm{\nabla}^2 S_N \cdot (\bm{q}-\Tilde{\bm{q}})/2\hbar} 
    \, \dd \bm{q}
    \nonumber\\
    &= \eu{-S_N(\Tilde{\bm{q}})/\hbar} \prod_{k=1}^{(N-1)f} \int \eu{-\lambda_k s_k^2/2\hbar} \, \dd s_k
    \nonumber\\
    &= \eu{-S_N(\Tilde{\bm{q}})/\hbar} \prod_{k=1}^{(N-1)f} \sqrt\frac{2\pi\hbar}{\lambda_k} 
    \nonumber\\
    &= \sqrt{\frac{(2\pi\hbar)^{(N-1)f}}{\det \bm{\nabla}^2 S_N}}\,\eu{-S_N(\Tilde{\bm{q}})/\hbar},\label{eq:samewell}
\end{align}
where $\lambda_k$ and $s_k$ are the eigenvalues and normal modes of $\bm{\nabla}^2 S_N(\tilde{\bm{q}})$.
In this way, the expression for the propagator %
that stays in a single well is given by
\begin{equation}
        \bra{q_{\mathrm{init}}}\eu{-\beta \hat{H}}\ket{q_{\mathrm{init}}} =  \lim_{N \rightarrow \infty}  A_N \sqrt{\frac{(2\pi \hbar)^{(N-1)f} }{(m\tau_N)^{(N-1)f}\det \bm{J}^{(0)}}} \, \eu{-S_N(\Tilde{\bm{q}}^{(0)})/\hbar},
\end{equation}
where in this case the minimum-action path, $\tilde{\bm{q}}^{(0)}$, has all beads located at the bottom of the well, $\tilde{q}_n^{(0)}=q_\mathrm{init}$, %
and we have introduced the notation $\bm{J}^{(0)}=\bm{\nabla}^2S_N(\tilde{\bm{q}}^{(0)})/m\tau_N$ to denote the mass-weighted Hessian of this collapsed polymer chain. %
From now on, we will assume that all quantities are evaluated in the $N\rightarrow\infty$ limit without explicitly specifying it.

\subsection{Instanton tunnelling path}

\begin{figure}
    \centering
    \includegraphics[width=0.5\linewidth]{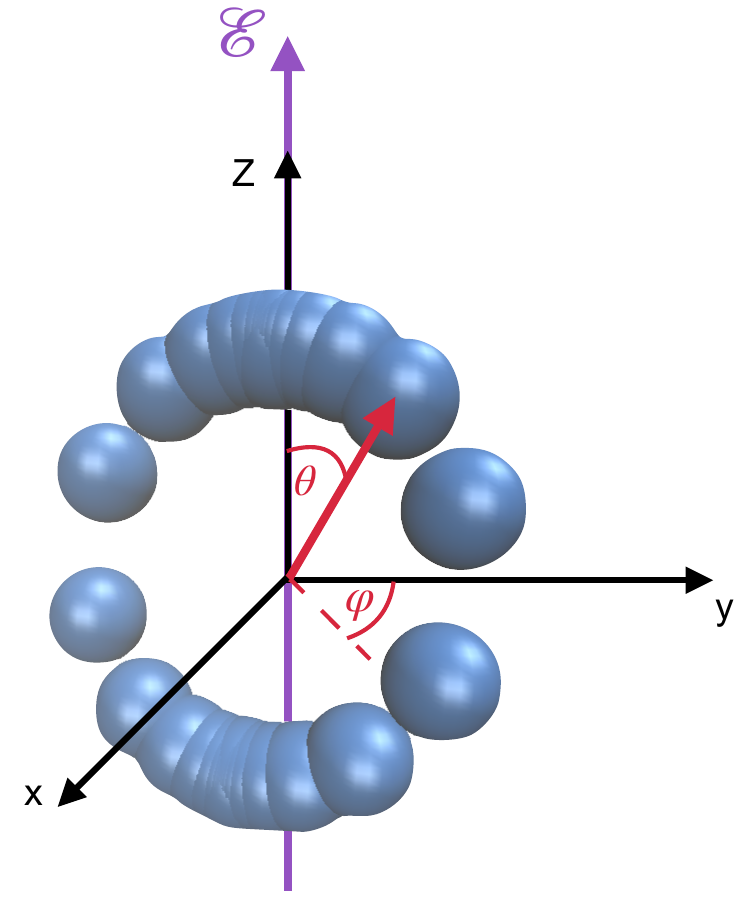}
    \caption{Illustration of one of the infinitely many instanton paths.
    The blue beads symbolize the positions of the hydrogen atoms along the tunnelling pathway (N.B.\ many more beads are used in practice to reach convergence).
    The higher density of beads at the top and bottom shows that the imaginary-time velocity of the atoms decreases as the molecule approaches its aligned configurations. Moreover, from this figure, it should be clear how the action is invariant to rotations of the instanton around the $z$-axis.}
    \label{fig:rot}
\end{figure}

The path associated with the reorientational tunnelling process corresponds to a rotation of the molecule from $\theta=0$ to $\theta=\pi$. %
The instanton is defined as the minimum-action pathway, $\tilde{\bm{q}}$, in full dimensionality and thus obeys the Euler--Lagrange equations in imaginary time, meaning that it is equivalent to a classical trajectory on the upside-down potential \cite{Miller1971density,coleman_uses_1979}.
Therefore, linear momentum must be conserved. %
In particular, as we choose the centre of mass of both $q_\mathrm{init}$ and $q_\mathrm{fin}$ to be equal, the conserved linear momentum is zero and the centre of mass stays fixed along the tunnelling pathway.
Additionally, the angular momentum about the $z$-axis is conserved due to the cylindrical symmetry of the problem, whereas the components about the $x$- and $y$-axes are not.
In fact, the minimum-action pathway has zero angular momentum about the $z$-axis, which ensures that the instanton path lies in a plane that includes the $z$-axis.
The resulting imaginary-time motion of the hydrogen atoms is illustrated in Fig.~\ref{fig:rot}.

The propagator which describes tunnelling between the two wells presents a subtle complication and cannot be evaluated directly via the approach used in Sec.~\ref{sec:semiclassical}. 
This difficulty arises due to the presence of two symmetries of the path integral, each of which gives rise to a zero-frequency mode. A direct application of Eq.~\eqref{eq:samewell} %
would thus lead to division by zero because two of the eigenvalues ($\lambda_1$ and $\lambda_2$) are zero.
This necessitates a separate treatment of the integrals over the zero modes.

\begin{figure}
    \centering
    \includegraphics[width=.5\linewidth]{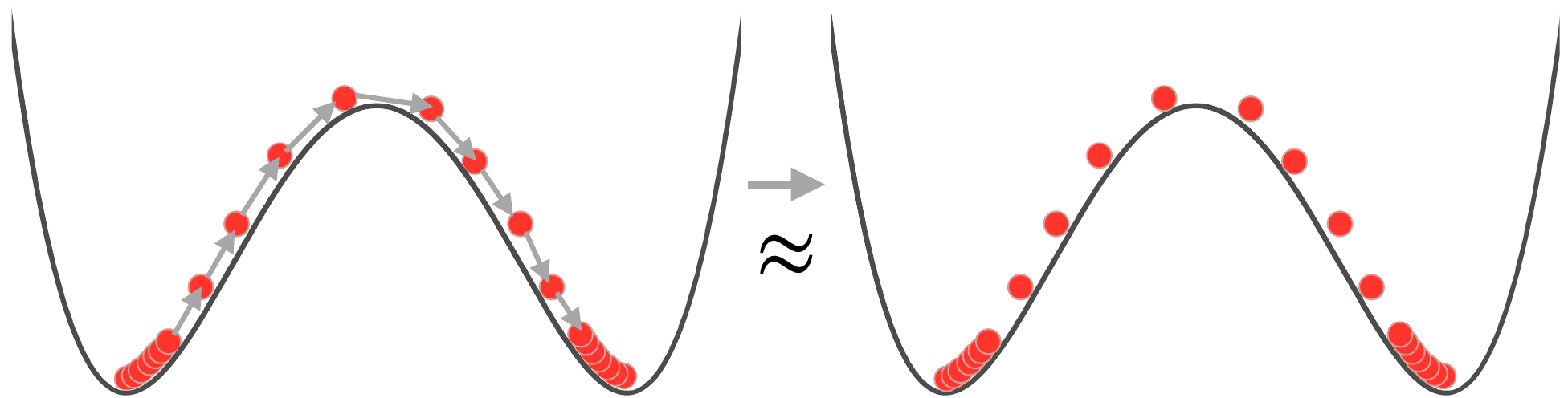}
    \caption{Illustration of the permutational zero mode. In the limit of an infinite number of beads at low temperature, there are infinitely many beads in each well.  Therefore, shifting each bead to the position of the next bead leaves the polymer effectively unchanged and hence the action invariant. This symmetry results in a zero-frequency mode of the polymer Hessian.}
    \label{fig:permutation}
\end{figure}

The first symmetry that gives rise to a zero-frequency mode originates from the permutation of beads within the polymer. 
In the $N\rightarrow\infty$ limit, %
the density of the beads increases until they
describe a continuous trajectory.
In addition, as the temperature is lowered towards zero, a large number of beads start to congregate at the bottom of the wells.
Consequently, shifting each position from $\tilde{q}_{n-1}$ to $\tilde{q}_{n}$, as illustrated in Fig.~\ref{fig:permutation}, leaves the action invariant and hence corresponds to a zero-frequency mode of the polymer Hessian.
Thankfully, there is a well-established procedure for treating this mode correctly \cite{Uses_of_Instantons,Benderskii,Kleinert,tunnel,AnharmInst}.
In particular, a normal-mode transform is applied to isolate the zero-frequency mode, and the integral over this mode is treated analytically.

There is, however, a second zero-frequency mode due to the special symmetry of this problem. Rotating the instanton around the axis of the electric field also leaves the action invariant, effectively generating infinitely many equivalent instanton pathways at different angles. %
Following the procedure developed to treat the permutational zero mode, we will describe a similar approach to treat the rotational zero mode.
Note that, in this case, there are no zero-frequency translational degrees of freedom because the ends of the polymer are fixed to the initial and final configurations.

\subsection{Permutational zero mode}
In the low-temperature and large-$N$ limits, the instanton exhibits time-translational invariance:
\begin{align}
    \rmd q_n = \dot{\tilde{q}}_n \, \rmd \tau = \frac{\tilde{q}_n-\tilde{q}_{n-1}}{\tau_N} \, \rmd \tau,
\end{align}
where $\dot{\Tilde{q}}_n$ %
is the mass-weighted imaginary-time velocity of the path at bead $n$.
Therefore the permutational zero mode $s_1$
is related to the original measure by 
\begin{equation} \label{eq:ds}
\mathrm{d}q_n
= \frac{1}{\sqrt{B_1}}  \, (\tilde{q}_{n}-\tilde{q}_{n-1}) \, \mathrm{d}s_1 .
\end{equation}
The transform has a normalization constant $B_1$ such that the Jacobian is unitary: %
\begin{equation}
    1 = \sum_n \left\lVert \frac{\partial q_n}{\partial s_1}\right\rVert^2
    = \frac{1}{B_1} \sum_n \lVert \Tilde{q}_n - \Tilde{q}_{n-1} \rVert^2 .
\end{equation}
It is thus clear that the normalization constant can be written as $B_1 = \sum_n \lVert \Tilde{q}_n - \Tilde{q}_{n-1} \rVert^2$, which is the mass-weighted squared path length.
However, we can find another expression for the normalization constant by recognizing that the instanton energy $E = -\frac{m}{2} \lVert \dot{\tilde{q}}\rVert^2 + V(\tilde{q})$ is constant and can be set to zero. Using this,
we can rewrite the instanton action %
as
\begin{equation}
    S_{\mathrm{inst}} = S_N(\tilde{\bm{q}}) = \int_{0}^{\beta\hbar} m \lVert \dot{\tilde{q}} \rVert^2 \, \dd \tau
    = \frac{m B_1}{\tau_N} . %
\end{equation}
The normalization constant is then given by $B_1 = S_{\mathrm{inst}}\tau_N/m$.
Finally, %
we can integrate over the zero-frequency mode to give
\begin{equation}
    \int \dd s_1 = \int_0^{\beta\hbar}\frac{\sqrt{B_1}}{\tau_N} \, \dd \tau %
    = \beta\hbar \sqrt{\frac{ S_{\mathrm{inst}}}{m\tau_N}}.
\end{equation}

\subsection{Rotational zero mode}
In addition to the well-known permutational zero mode,
our problem has a second zero mode resulting from rotational symmetry.
The symmetry comes from the fact that rotating the instanton path around the axis of the electric field leaves the action invariant, effectively yielding infinitely many equivalent instanton pathways with different values of $\varphi$. This is illustrated in Fig.~\ref{fig:rot}.
It is important to realize that this is not a symmetry of the molecule itself, but of the instanton path. 
In fact, the rotational symmetry is not present in the collapsed path where \ce{H2} is static and aligned with the field.
Like all linear molecules, it has no degree of freedom associated with rotation around the $z$-axis.
In summary, we have a rotational symmetry of the path that arises in addition to the other symmetries of the molecule.
In the following, we will describe how one can analytically integrate over the rotational zero mode, in a manner similar to the permutational mode discussed in the previous section.

First, we must consider how the %
configurations of each bead change
when the instanton is rotated around the axis of the electric field. %
A rotation of the instanton moves each atom of each bead from $\tilde{q}_{n,a}=(\tilde{x}_{n,a},\tilde{y}_{n,a},\tilde{z}_{n,a})$ to $q_{n,a}(\varphi) = R_z (\varphi) \tilde{q}_{n,a}$,
where the rotation matrix is
\begin{equation}
R_z(\varphi) =
\begin{bmatrix}
\cos\varphi & -\sin\varphi & 0  \\
\sin\varphi & \phantom{+}\cos\varphi  & 0  \\
0          & \phantom{+}0           & 1  \\
\end{bmatrix} .
\end{equation}
We can now differentiate with respect to $\varphi$ and set $\varphi=0$ to obtain the tangent vector
\begin{equation}
    \frac{\partial q_{n,a}}{\partial \varphi} = \frac{\partial R_z}{\partial \varphi}\biggl|_{\varphi = 0} \tilde{q}_{n,a} = \begin{pmatrix}-\tilde{y}_{n,a} \\ +\tilde{x}_{n,a}\\ 0\end{pmatrix} . \label{eq:tangent}
\end{equation}

The rotational zero mode, $s_2$, is then
related to the original integration variables by 
\begin{equation} \label{eq:s2}
    \dd q_n
    = \frac{\partial q_n}{\partial \varphi} \, \dd \varphi
    = \frac{1}{\sqrt{B_2}} \frac{\partial q_n}{\partial \varphi} \, \dd s_2 ,
\end{equation}
with the normalization condition
\begin{equation}
    1 = \sum_n \norm{\frac{\partial q_n}{\partial s_2}}^2 =
    \frac{1}{B_2} \sum_n \norm{\frac{\partial q_n}{\partial \varphi}}^2 .\label{eq:norm}
\end{equation} %
From this and Eq.~\eqref{eq:tangent}, 
the normalization constant is seen to be
\begin{align}
B_2 =\sum_n \sum_a  (\tilde{x}_{n,a}^2 + \tilde{y}_{n,a}^2) = \frac{I_{zz}}{m} ,
\end{align}
where the index $a$ sums over the atoms and the index $n$ sums over the beads in the polymer.
We notice that the normalization constant is proportional to $I_{zz}$, the moment of inertia of the whole instanton around the $z$-axis.
Finally, using Eq.~\eqref{eq:s2}, the integral over the rotational zero mode can be rearranged to give
\begin{equation}
 \int \mathrm{d}s_2 =  \int_0^{2\pi} \sqrt\frac{I_{zz}}{m}  \, \dd \varphi = 2 \pi \sqrt\frac{I_{zz}}{m} .
\end{equation}

\subsection{Final expression}
We can now evaluate the propagator between the two wells as the large-$N$ limit of 
\begin{align}
    \bra{q_{\mathrm{fin}}}&\mathrm{e}^{-\beta \hat{H}}\ket{q_{\mathrm{init}}} \simeq A_N \cdot 2 \pi \sqrt\frac{I_{zz}}{m} \cdot \beta\hbar \sqrt{\frac{ S_{\mathrm{inst}}}{\tau_N m}} \cdot \sqrt{\frac{(2\pi \hbar)^{(N-1)f-2} }{(m\tau_N)^{(N-1)f-2}\det'' \bm{J}}} \, \eu{-S_{\mathrm{inst}}/\hbar} ,
\end{align}
where $\bm{J}=\bm{\nabla}^2S_N(\tilde{\bm{q}})/m\tau_N$ is the mass-weighted Hessian matrix of the fixed-ended polymer potential for the propagator between the two wells. 
Note that two zero-frequency modes are excluded from the determinant of $\bm{J}$ as indicated by the double prime.

Finally, we can determine the tunnelling splitting by evaluating the ratio of the propagators as described above in Eq.~\eqref{eq:Deltafromprop}. Putting everything together, we obtain:\footnote{Note that we have additionally used the asymptotic relation $\arctanh(x) \simeq x$, which is valid in the semiclassical limit \cite{AnharmInst}.}
\begin{align}
    \Delta 
    &\simeq \lim_{\beta \rightarrow \infty}  \frac{2}{\beta} \frac{\bra{q_{\mathrm{fin}}}\mathrm{e}^{-\beta \hat{H}}\ket{q_{\mathrm{init}}}}{\bra{q_{\mathrm{init}}}\mathrm{e}^{-\beta \hat{H}}\ket{q_{\mathrm{init}}}} \nonumber %
    \\&\simeq\lim_{\beta \rightarrow \infty}2\sqrt{\tau_N I_{zz}  S_{\mathrm{inst}}}\cdot \sqrt{\frac{\det \bm{J}^{(0)}}{\det''\bm{J}}} \, \eu{-S_{\mathrm{inst}}/\hbar} .
    \label{eq:finalts}
\end{align}
This formula is the final working equation of our semiclassical instanton theory for calculating the tunnelling splitting of a molecule in an electric field, or more generally, for any problem where the instanton has a rotational symmetry axis not found in the equilibrium geometry.

\section{Methods and Results}
In the previous section, we established the theoretical framework to address the problem of a molecule tunnelling in an electric field. Our next step is to validate the derived equations by applying them to a simple case such as the hydrogen molecule, where alternative methods for calculating the tunnelling splitting are available.
As discussed in the introduction, the tunnelling splitting of \ce{H2} in an electric field corresponds to its ortho--para splitting.  A static electric field will align the molecule.  However, since it does not have a permanent dipole moment, there is no preference for its orientation, leading to two potential wells of equal depth as shown in Fig.~\ref{fig:ts}.

To benchmark our theory, we use perturbation theory in the electric-field strength to develop a model potential similar to those employed by Friedrich and Herschbach \cite{friedrich_polarization_1995} for which we can evaluate the quantum-mechanical tunnelling splitting and compare with the instanton result.
Next, we move on to a non-perturbative treatment of the field by including it explicitly in accurate electronic-structure calculations evaluated ``on the fly'' while optimizing the instanton. %
After this, we compare the results of the perturbative approach to the non-perturbative treatment of the electric field, and discuss the impact of such approximations. %

Finally, we study a polar molecule, formaldehyde, under intense non-resonant laser radiation, which is essentially a strong electric field oscillating at high frequency.
In a typical case, the field oscillates so fast %
that the interaction with the molecule can be averaged over an oscillation period %
\cite{friedrich1999manipulating}. This leads to an effective time-independent potential energy surface that is qualitatively similar to that of a non-polar molecule in a static field, with two potential wells of equal depth between which the system can tunnel. We apply instanton theory to this problem in Sec.~\ref{sec:formaldehyde}. %

\subsection{H$_2$ -- perturbative model} \label{sec:H2model}
\begin{figure}
    \centering
    \includegraphics[width=0.49\linewidth]{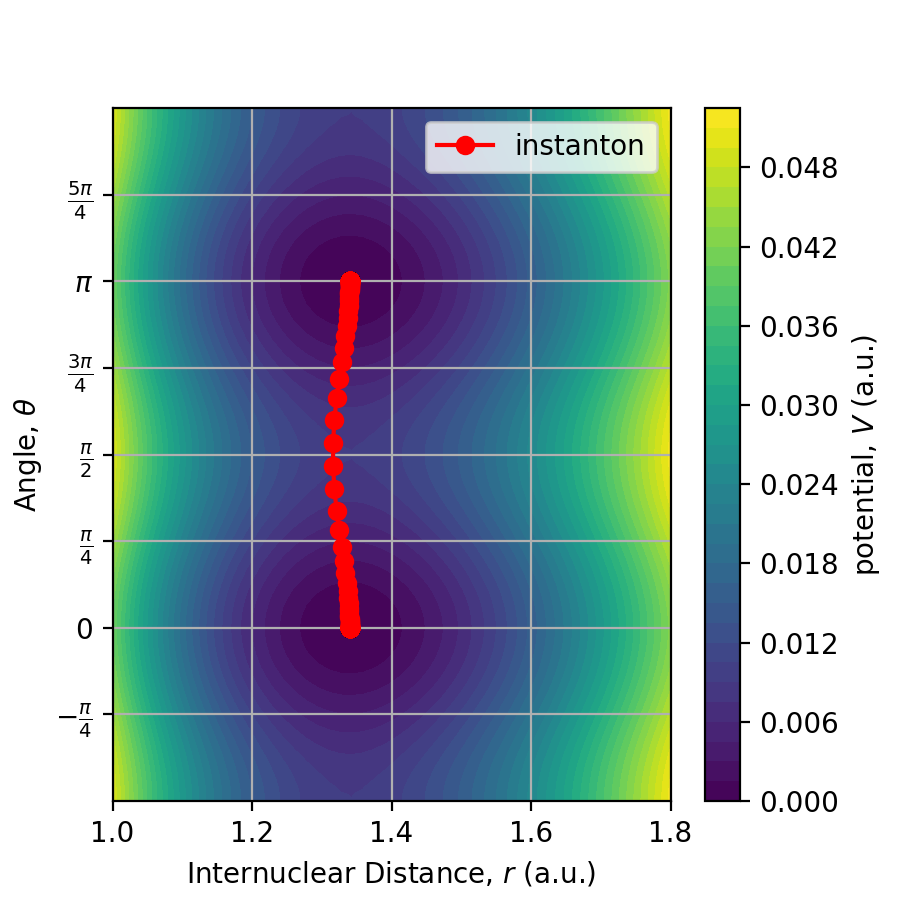}
    \caption{Plot of the model potential energy surface defined in Eq.~\eqref{eq:pot}. %
    There are two minima corresponding to the aligned orientations of the hydrogen molecule.
    The optimized instanton pathway is shown in red. %
    One can see that the bond length decreases slightly during the tunnelling process---similar to an ice skater pulling in their arms while performing a spin.}
    \label{fig:instonpot}
\end{figure}

First, we construct a simple model of \ce{H2} in an electric field.
We define the field to be along the direction of the $z$-axis of a fixed coordinate frame. %
Following Friedrich and Herschbach \cite{friedrich_alignment_1995},
we express the interaction energy as
\begin{equation}
    V_\text{int}(\theta) = \tfrac{1}{2}(\alpha_\perp \cos^2\theta + \alpha_\parallel \sin^2\theta)\mathscr{E}^2 ,
\end{equation} 
where $\alpha_\perp$ and $\alpha_\parallel$ are the polarizability orthogonal and parallel to the intramolecular axis, respectively,
which we set
to the experimentally determined values %
$\alpha_\perp=4.57769$ a.u.\ and $\alpha_\parallel=6.38049$ a.u. \cite{kolos_polarizability_1967}.
However, we go beyond the one-dimensional treatment of Ref.~\cite{friedrich_alignment_1995} by additionally accounting for the vibrational degree of freedom
by a Morse oscillator: %
\begin{equation}
    V(r, \theta) = D_\mathrm{e} (1- \mathrm{e}^{-a(r-r_{\mathrm{eq}})}) -V_{\mathrm{int}}(\theta) , \label{eq:pot}
\end{equation}
with the equilibrium bond length $r_{\mathrm{eq}} = 1.34$ a.u.,
well depth $D_\mathrm{e} = 0.56$ a.u., and the width of the potential $a = \sqrt{\frac{k}{2D_\mathrm{e}}} = 0.7$ a.u., where $k$ is the force constant of the bond.
The resulting potential energy surface is shown in Fig.~\ref{fig:instonpot}.

Using this potential energy  surface, we can now calculate the tunnelling splitting by optimizing the instanton and evaluating Eq.~\eqref{eq:finalts}.
The number of beads $N$ and inverse temperature $\beta$ are increased until %
the results converge to the reported precision. %
We have repeated this process for two different field strengths, $\mathscr{E}=0.08$\,a.u.\ $\approx 4.1\cdot10^{8}$\,V/cm and $\mathscr{E}=0.1$\,a.u.\ $\approx 5.1\cdot10^{8}$\,V/cm.

In Table~\ref{tab:compare}, we compare the tunnelling splittings predicted by instanton theory with the quantum-mechanical splittings obtained by numerically solving the Schrödinger equation (as described in Appendix ~\ref{app:DVR}).
The instanton results are in good agreement with the benchmark.
In both cases, the tunnelling splitting is strongly dependent on the field strength and decreases sharply as the field is increased.
This is because stronger fields lead to larger barrier heights such that the tunnelling is inhibited.
Note that the semiclassical approximation underlying instanton theory is most accurate in cases where the tunnelling splitting is small and the barrier is large \cite{tunnel,richardson_ring-polymer_2018,AnharmInst}.  For this reason, the relative error of instanton theory is observed to decrease when increasing the field strength and we expect it to tend exactly to the quantum result in the limit of an infinite field.

\begin{table}[h]
    \caption{Comparison between tunnelling splittings calculated with quantum mechanics (QM) and semiclassical instanton theory for two electric-field strengths. %
    }
    \centering
    \renewcommand{\arraystretch}{1.2} %
    \setlength{\tabcolsep}{5pt}       %
    \arrayrulewidth=1pt               %
    \begin{tabular}{cccc}
       \hline
        $\mathscr{E}$ (a.u.) & $
        \Delta_{\mathrm{QM}}$ (cm$^{-1}$) & $\Delta_\mathrm{inst}$  (cm$^{-1}$)  & deviation \\
        \hline
                {0.08} & 5.03  & 6.61 & 31\%  \\
        {0.1\phantom{0}}  & 0.97   & 1.19  & 23\%  \\
        \hline
    \end{tabular}
    \label{tab:compare}
\end{table}

In addition to predicting the energy splitting, instanton theory provides mechanistic insight into the tunnelling process.
Figure~\ref{fig:instonpot} shows that as the molecule rotates, the internuclear distance decreases slightly.
At first sight, this behaviour may appear counterintuitive, as one might expect the centrifugal force to elongate the bond.
However, an increased internuclear distance would increase the moment of inertia of the molecule making rotation more difficult.
Therefore, in order to minimize the action, the optimal tunnelling pathway chooses to decrease the internuclear distance. 
This motion can be compared to an ice skater tugging in their arms during spins to increase their speed; the instanton uses this same concept to 
maximize its tunnelling probability.

Overall, our results for this test model validate our generalization of the semiclassical instanton theory to this problem.
We next turn to applications to more complex systems where other theories are unavailable or impractical to use.

\subsection{H$_2$ -- non-perturbative \textit{ab initio} treatment}
In the previous section, %
we made several simplifying assumptions in order to develop a simple model of H$_2$ in an electric field: (i) we approximated the (field-free) attraction between the atoms by a Morse potential; (ii) we assumed the field strength to be sufficiently weak such that we could approximate the interaction with the electric field only using the polarizability of the H$_2$ molecule, ignoring hyperpolarizabilities and other higher-order terms
\cite{stone_theory_2013}; (iii) we used the polarizabilities of H$_2$ in its equilibrium geometry, and assumed them to be independent of the bond length. In this section, we lift all of these assumptions by including the electric field explicitly in our electronic-structure calculations. %
A major advantage of instanton theory is that knowledge of only a small region on the PES is needed, allowing us to perform high-level electronic-structure calculations ``on the fly" without needing to construct a global fit for the PES.

To obtain highly accurate electronic structure data, we employed the coupled-cluster singles and doubles (CCSD) method in conjunction with the augmented correlation-consistent polarized valence triple-zeta (aug-cc-pVTZ) basis set.\footnote{Increasing the size of the basis set to aug-cc-pVQZ reduced the action $S_\mathrm{inst}/\hbar$ from 10.57 to 10.55, which is only expected to reduce the predicted tunnelling splitting by about 2\%.} 
Note that in the limit of an infinitely large basis set, CCSD is exact for H$_2$, as there are only two electrons present.
The external static electric field was applied along the $z$-axis using the command \textsf{efield} in ORCA 5 \cite{neese_orca_2012, neese_software_2022}.
Analytic gradients were available but Hessians were obtained using finite differences.

In order to facilitate a comparison between this problem and the model system from the previous section, geometry optimizations were carried out, in particular to determine the equilibrium internuclear distance and the barrier height (defined as the difference in energy between the relaxed molecular geometry aligned with and perpendicular to the field). The results are shown in Table~\ref{tab:H2nonpert}. 
The first thing to note is that the \textit{ab initio} calculations predict a significant elongation of the equilibrium bond length under an applied electric field.
Additionally, as shown in Fig.~\ref{fig:instonpot}, we expect a further change in the bond length along the instanton path.
Both of these observations suggest that assuming a constant polarizability fixed at its field-free value may not be a valid approximation.
In fact, this explains the observed deviation of the barrier height, $V^{\ddag}$, between the model PES and the \textit{ab initio} calculations. %
Note that the effect of including the second hyperpolarizability \cite{bishop_theoretical_1991} is only expected to increase the barrier height by about 5\% for $\mathscr{E}=0.1$\,a.u,\footnote{Using the values from Ref.~\cite{bishop_theoretical_1991} in atomic units, the correction is $(682.5-575.9)\mathscr{E}^4/4!=97.5\,\mathrm{cm}^{-1}$} implying that the main error of the model PES comes from the assumption of a constant polarizability rather than the neglect of higher-order terms.

\begin{table}
\caption{Comparison of the instanton results for \ce{H2} in an electric field using the perturbative model and the \textit{ab initio} calculations.}
\label{tab:H2nonpert}
\centering
\renewcommand{\arraystretch}{1.3}
\arrayrulewidth=1pt               %
\begin{tabular}{@{}lcccccc@{}}
\hline
 &$\mathscr{E}$ (a.u.)& $r_\mathrm{eq}$ (a.u.) & $V^\ddag$ (cm$^{-1}$) & $\Delta$ (cm$^{-1}$) \\
\hline
\multirow{2}{*}{model PES}
  & $0.08$ & 1.34   & 1266  & 6.61 \\
  & $0.1\phantom{0}$  & 1.34   & 1978  & 1.19 \\
\hline

\multirow{2}{*}{\textit{ab initio}}
  & $0.08$ & 1.4831 & 1614  & 1.5 \\
    & $0.1\phantom{0}$  & 1.5873 & 3290 & -- \\
\hline
\end{tabular}

\end{table}

Next, we optimized a fixed-ended polymer chain for $\mathscr{E}=0.08$\,a.u.\ by minimizing its action, while constraining the endpoints to the %
two wells optimized under the same field.
The resulting instanton path was then used to compute the tunnelling splitting of the molecule using Eq.~\eqref{eq:finalts}.
Note that, we did not compute the instanton for the case of $\mathscr{E}=0.01$\,a.u.\ as the tunnelling splitting is expected to be very small due to the high barrier.
Table~\ref{tab:H2nonpert} gives a comparison of the tunnelling splitting predicted for the model PES with the \textit{ab initio} calculations.
The \textit{ab initio} result is almost 5 times smaller, which highlights the considerable error caused by the perturbative approximation.

\subsection{Formaldehyde} \label{sec:formaldehyde}
So far, we have demonstrated how to apply instanton theory to a simple, non-polar diatomic molecule in a static electric field. We now increase the complexity of the problem by examining formaldehyde (\ce{H_2CO}), a polar, non-linear polyatomic molecule in full dimensionality. %
In a static electric field, a polar molecule would not only align with the field, but also preferentially orient itself in the field according to its dipole. The two orientations are not degenerate and no tunnelling splitting occurs. However, if we instead place the molecule into an electric field created by a non-resonant laser, it is possible to %
restore the degeneracy.
For optical frequencies, the timescale of the oscillations in the electric field is typically very small compared to the tunnelling process \cite{friedrich_polarization_1995}.  If we additionally assume that the laser pulse is long compared to the oscillation period, it is justified to average over the oscillations.
This effectively cancels all terms containing odd powers of the electric field (such as the electric-dipole interaction); only even terms remain, so that the field can only induce alignment and not orientation. 

The laser is modelled by an oscillating electric field in the $z$-direction with amplitude $\mathscr{E}(t) = \mathscr{E}_0 \cos \omega t$. %
After averaging, the effective multidimensional potential energy surface felt by the molecule is thus
\begin{subequations}
\begin{align}
    \langle V\rangle
    &= \left\langle V_0 - \mu_z\mathscr{E}(t) - \frac{1}{2} \alpha_{zz}\mathscr{E}(t)^2 - \frac{1}{3!} \beta_{zzz}\mathscr{E}(t)^3 -  \frac{1}{4!} \gamma_{zzzz}\mathscr{E}(t)^4 - \cdots \right\rangle \\
    &= V_0 - \frac{1}{4} \alpha_{zz}\mathscr{E}_0^2 - \frac{1}{64} \gamma_{zzzz} \mathscr{E}_0^4 - \cdots,\label{eq:averagedpot}
\end{align}
\end{subequations}
where we have defined $V_0(q)$ as the Born--Oppenheimer potential without the contribution of the electric field and $\mu_z$, $\alpha_{zz}$, $\beta_{zzz}$ and $\gamma_{zzzz}$ are the $q$-dependent dipole, polarizability and hyperpolarizabilities in the direction of the electric field \cite{stone_theory_2013}.
Moreover, we have used $(2\pi/\omega)^{-1} \int_0^{2\pi/\omega}\cos^2(\omega t) \,\dd t = 1/2$ and %
$(2\pi/\omega)^{-1} \int_0^{2\pi/\omega}\cos^4(\omega t) \,\dd t = 3/8$, as well as the fact that all odd-terms average to zero.
The effect of the field is most easily included in our \textit{ab initio} methodology by performing two electronic-structure calculations, one with a static field pointing `up', $\mathscr{E}_0/\sqrt{2}$, and one pointing `down', $-\mathscr{E}_0/\sqrt{2}$, which are then averaged. This corresponds to 
\begin{equation}
   V = \frac{V_{\uparrow} + V_{\downarrow}}{2} = V_0 - \frac{1}{4} \alpha_{zz}{\mathscr{E}}_0^2 - \frac{1}{32} \gamma_{zzzz}{\mathscr{E}}_0^4 - \cdots ,
\end{equation}
which agrees with the time-averaged laser field [Eq.~\eqref{eq:averagedpot}] up to second order in $\mathscr{E}_0$.\footnote{In this case (where we use weaker fields than for \ce{H2}), effects from hyperpolarizabilities %
are expected to be negligible especially in comparison to the intrinsic errors of the density-functional approximation.}

We performed density functional theory (DFT) calculations (with an explicit electric field) using the B3LYP hybrid exchange--correlation functional and the def2-TZVPP triple-zeta valence basis set with double polarization \cite{weigend_balanced_2005}, as implemented in ORCA 6 \cite{neese_software_2025}.
These calculations were unfortunately prone to numerical instabilities and the numerical Hessian routine in ORCA was not able to calculate the second derivatives with sufficient accuracy, in particular for weak fields where the eigenvalue spectrum covers a wide range from the high-frequency C--H bonds to the low-frequency pendulum motions.
Thus, we employed Ridders' method \cite{NumRep} to calculate numerical derivatives of the analytical gradients provided by ORCA.
This technique enhances the accuracy of the finite-difference approximation by computing the derivative with multiple step sizes %
and extrapolating \cite{richardson_ix_1911} to the limit of an infinitesimal step.

Moreover, in order to enhance the convergence of the instanton calculations, %
we generated initial guesses using the results of previous smaller instanton optimizations.
In particular, when increasing the number of beads, $N$, we simultaneously scaled the inverse temperature, $\beta$, by the same factor (\emph{e.g.} a doubling of $N$ is accompanied by a doubling in $\beta$). This strategy enables us to generate a new guess by introducing additional beads at the ends rather than by introducing new beads along the path, %
which significantly reduces the number of optimization steps required at each temperature.

\begin{figure}
    \centering
    \includegraphics[width=.4\linewidth]{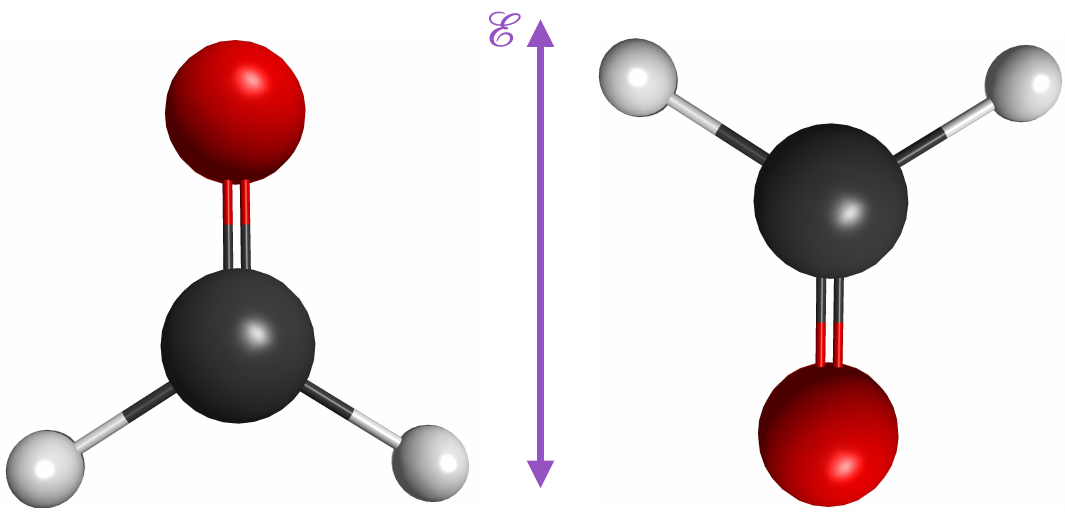}
    \caption{Illustration of the two minimum-energy orientations of the formaldehyde molecule in an oscillating laser field. The purple line represents the axis of the electric field with the arrows in both directions symbolizing that the molecule feels no directionality of the field.}
    \label{fig:form}
\end{figure}

\begin{figure}
    \centering
    \includegraphics[width=0.5\linewidth]{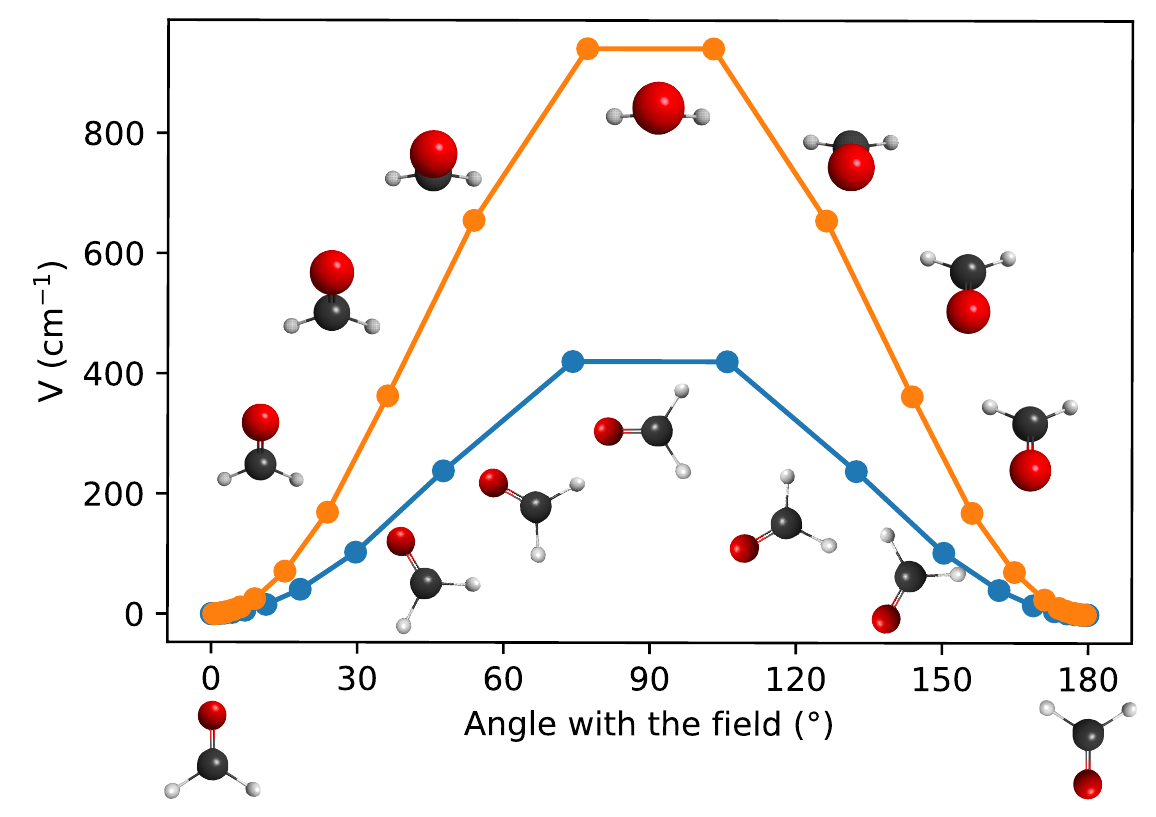}
    \caption{Comparison of the instanton tunnelling paths corresponding to the two possible rotations that the \ce{H2CO} molecule can perform. In both cases, the potential at a field strength of $\mathscr{E}=0.03$ a.u.\ is plotted against the angle between the C=O bond and the field. The instantons shown were obtained with $\beta=40\,000$\,a.u. and $N=32$.
    The rotation in the plane of the molecule (blue) has a lower potential barrier %
    and is thus strongly favoured over the out-of-plane rotation (orange).}
    \label{fig:barrier}
\end{figure}

With this framework, we were able to apply instanton theory to calculate the tunnelling splitting of formaldehyde in an electric field. The two minimum-energy orientations of the molecule are shown in Fig.~\ref{fig:form}.
It is important to note that the molecule can undergo two distinct tunnelling pathways when reorienting from the oxygen-up to the hydrogen-up configuration: (i) a rotation within the molecular plane; and (ii) a rotation out of this plane.
Comparing the potential energy along the two instanton paths as shown in Fig.~\ref{fig:barrier} indicates that the in-plane rotation provides the energetically preferred tunnelling pathway.
Since the tunnelling splitting depends exponentially on the action (which in turn increases with a larger barrier), the contribution from the higher-action pathway is subdominant.\footnote{See related work in which trans and cis tunnelling motions were compared for molecules like \ce{H2O2} \cite{chiral}.} We can therefore safely restrict our analysis to the in-plane mechanism without loss of accuracy.

Analysing the symmetry, we notice that there is no rotational invariance for either the collapsed or the tunnelling path, because the ends of the polymer chain are fixed in a particular configuration.
This is an important difference from the case of the linear \ce{H2} molecule, which means that we do not need to integrate out the rotational zero mode explicitly when using fixed-ended path integrals for formaldehyde. We thus calculate the tunnelling splitting using
\begin{equation}
        \Delta=\lim_{\beta \rightarrow \infty}2\sqrt{\tau_NS_{\mathrm{inst}}}\cdot \sqrt{\frac{\det \bm{J}^{(0)}}{\det'\bm{J}}} \, \eu{-S_{\mathrm{inst}}/\hbar},\label{eq:form}
\end{equation}
where we have only removed the permutational zero mode from the determinant of $\bm{J}$.\footnote{Note that, the eigenvalues corresponding to the lowest-frequency translational and rotational fluctuations were very small and could be numerically unstable. In practice, we thus discarded the lowest 4 eigenvalues of $\bm{J}_0$ and the lowest 5 eigenvalues of $\bm{J}$.}

We applied electric fields of $\mathscr{E}_0=0.01$\,a.u.\ $\approx 5.1\cdot10^7$\,V/cm, $\mathscr{E}_0=0.03$\,a.u.\ $\approx 1.5\cdot10^8$\,V/cm and $\mathscr{E}_0=0.05$\,a.u.\ $\approx 2.6\cdot10^{8}$\,V/cm, which are within the current experimental limit for the fields induced by high-intensity lasers \cite{boll_imaging_2014}.
Using field strengths of $\mathscr{E}_0=0.03$\,a.u.\ and 0.05\,a.u., we compute tunnelling splittings of the order of $10^{-14}$\,cm$^{-1}$ and $10^{-23}$\,cm$^{-1}$.
These extremely small tunnelling splittings are unlikely to be experimentally observable.
It should, however, be possible to significantly increase the tunnelling splitting by reducing the field strength.
Unfortunately, obtaining Hessians with the required accuracy was not possible with the current protocol for the electric-field strength of $\mathscr{E}_0=0.01$\,a.u., even when employing Ridders' method.
However, it is possible to optimize an instanton for this case.
We can then use this result to estimate the tunnelling splitting by calculating the action and multiplying this with the pre-exponential factor of Eq.~\eqref{eq:form} as calculated for the other two electric-field strengths.
This is possible because a comparison of the computed values for this prefactor at higher field strengths shows that it remains roughly constant and we assume this behaviour to continue giving us an estimated result for the tunnelling splitting at a lower field strength.
These calculations are displayed in Table \ref{tab:tsform_transposed} where the estimated result at the lowest field strength is on the order of 10$^{-3}$\,cm$^{-1}$ and thus accessible to high-resolution spectroscopy.
Note that this splitting is a result of large-amplitude heavy-atom tunnelling, and yet is no smaller than splittings observed in water clusters due to hydrogen tunnelling \cite{hexamerprism}.
The reason for this is that in the present case, we can tune the electric-field strength to lower the barrier in order to compensate for the heavier mass.

\begin{table}[b]
\caption{Instanton calculations of the tunnelling splitting of \ce{H2CO} at various electric-field strengths.
The prefactor refers to the pre-exponential term of Eq.~\eqref{eq:form}.
Estimates based on the assumption of a constant prefactor are shown in parentheses.}
\label{tab:tsform_transposed}
\centering
\renewcommand{\arraystretch}{1.3}
\arrayrulewidth=1pt               %
\begin{tabular}{cccc}
\hline
$ \mathscr{E}_0$ (a.u.)& $S_\mathrm{inst}/\hbar$ & prefactor (a.u.) & $\Delta$ (cm$^{-1}$) \\
\hline
$0.01$ & 13.55 & ($5\cdot10^{-3}$) & $(1\cdot10^{-3})$ \\
$0.03$ & 39.11 & $5\cdot10^{-3}$ & $1\cdot10^{-14}$ \\
$0.05$ & 59\phantom{.00} & $4\cdot10^{-3}$ & $2\cdot10^{-23}$ \\
\hline
\end{tabular}

\end{table}

\begin{figure}[t!]
    \centering
    \includegraphics[width=0.5\linewidth]{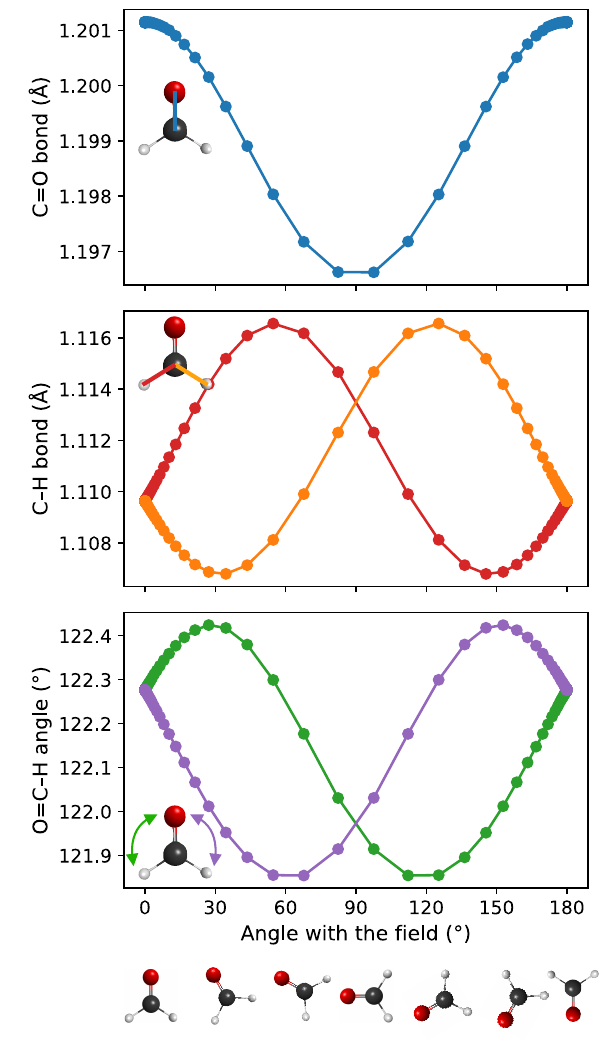}
    \caption{Changes of the molecular geometry of \ce{H2CO} along the tunnelling path. As the molecule rotates, we observe a shortening of the C=O bond similar to the example of \ce{H2}. 
    Moreover, the hydrogen atoms behave in an asymmetric way analogous to a cartwheel motion, which can be observed in both the C--H bond lengths as well as the O=C--H angles.}
    \label{fig:cartwheel}
\end{figure}

Furthermore, instanton theory allows us to analyse the motion of the molecule along the tunnelling path.
Figure~\ref{fig:cartwheel} reveals that, similar to the hydrogen molecule, the C=O bond shortens during the rotation to reduce the moment of inertia. %
In addition, we observe that the C--H bond lengths and O=C--H angles change in an asymmetric manner. This can be understood by comparing the molecular rotation %
to an acrobat performing a cartwheel: in that motion, %
the body does not remain rigid but limbs move in a complex rhythm,
creating an analogous pattern of asymmetry during rotation.

This molecular cartwheel motion is not only conceptually interesting but also has qualitative implications on the calculation since bond length and angles influence the polarizability of the molecule. 
In fact, it suggests that assuming a rigid geometry during the tunnelling motion would lead to inaccurate results. Instanton theory thus provides an advantage over other methods which cannot account for these changes.

\section{Conclusions}
In this work, we have employed semiclassical instanton theory to calculate tunnelling splittings induced by strong electric fields in both non-polar and polar molecules. %
For this, it was necessary to generalize the theory to account for the rotational symmetry of the instanton pathway in addition to the usual permutational mode.
We validated our approach by benchmarking the results for a model system representing a hydrogen molecule in an electric field that we treat perturbatively. 

In addition to predicting the tunnelling splitting, instanton theory also provides a picture of the mechanism of the process.
In particular, while the molecule is rotating to flip its orientation, the bond length was found to shrink slightly so as to reduce the moment of inertia. %
The tunnelling mechanism thus differs from a rigid rotation and instead follows a multidimensional pathway. %
Following previous studies \cite{friedrich_polarization_1995}, our model system set the value of the polarizability to its experimental value measured in equilibrium under zero-field conditions.
However, we show that this approximation is no longer valid, as the polarizability should be dependent on the bond length which changes along the tunnelling pathway.
We have therefore gone beyond the perturbative approximation by including the electric field explicitly within the electronic-structure calculations. As instanton theory allows for an ``on-the-fly" evaluation of the PES, it can be easily combined with \emph{ab initio} calculations at a high level of theory, yielding accurate estimates for the tunnelling splitting in full dimensionality.
We find qualitative and quantitative differences with the perturbative model.  In particular, there is an increase of the equilibrium bond length in a strong electric field, which leads to a higher barrier and hence suppressed tunnelling.

Additionally, we have demonstrated the power of instanton theory in going beyond simple diatomic molecules by calculating the tunnelling splitting of formaldehyde, a polar, nonlinear polyatomic molecule, in an oscillating laser field.
This molecule carries out an asymmetric cartwheel motion to flip between its two degenerate orientations with a frequency that strongly depends on the field strength.
With this example, we show that even %
heavy-atom tunnelling can lead to observable splittings,
posing a challenge for spectroscopy to %
e{probe these molecular gymnastics and
confirm the theoretical predictions.
If this setup can be realized in a laboratory, it offers the possibility to tune the barrier height of the potential by changing the electric-field strength, providing a means to systematically test instanton theory (and other quantum tunnelling methods) against experiment in various limits.

Although we have focused on semiclassical instanton theory in this work, other computational methods of higher accuracy could also be developed for this case, either by including anharmonic perturbative corrections \cite{AnharmInst} or using exact path-integral sampling \cite{Matyus2016tunnel2,PIMDtunnel,malonaldehydePIMD,Zupan2025}.
The semiclassical theory could also be extended to treat slightly asymmetric systems \cite{asymtunnel,fiechter_ring-polymer_2025} and vibrationally excited states \cite{Milnikov2005,Erakovic2020excited,laude_instanton_2024}.
In a forthcoming publication \cite{MagneticInst}, we will describe a further extension of the theory for molecules tunnelling in strong magnetic fields, such as those observed in neutron stars \cite{schmelcher1997molecules,monzel_molecular_2022,zalialiutdinov_exploring_2025}.

\section*{Acknowledgements}
This work is dedicated to Fr\'ed\'eric Merkt on the occasion of his 60th birthday,
a wonderful mentor, colleague and inspiring scientist.
J.A.G. thanks the Günthard Foundation and the German Academic Scholarship Foundation which support her undergraduate studies.
M.R.F. was supported by an ETH Grant.
\bibliographystyle{tfo}
\bibliography{amirasreferences,references,marit,other}

\appendix
\section{Quantum-mechanical tunnelling splitting calculations} \label{app:DVR}
To validate our newly derived instanton theory, we benchmarked the tunnelling splittings against fully quantum-mechanical calculations.
This comparison was carried out using the perturbative model of \ce{H2} from Sec.~\ref{sec:H2model}.
Whereas the instanton is evaluated in Cartesians,
the quantum calculations employed 3D spherical coordinates $(r, \theta, \varphi)$ to describe the hydrogen molecule in the electric field (with the centre-of-mass motion removed).
Importantly, the potential energy surface given in Eq.~\eqref{eq:pot} does not depend explicitly on the azimuthal angle $\varphi$. This cylindrical symmetry allows us to factorize the eigenfunctions in terms of a 2D wavefunction for the radial $r$ and polar $\theta$ coordinates, %
and a simple dependence on
the azimuthal angle $\varphi$:
\begin{equation}
    \Psi_m(r, \theta, \varphi) = r^{-1} \, \chi_m (r, \theta) \, \eu{i m\varphi},
\end{equation}
where $m$ denotes the azimuthal quantum number. %

The resulting 2D eigenvalue problem with reduced mass $\mu=m_1m_2/(m_1+m_2)$ is
\begin{align}
 \hat{H}_m \chi_m =& -\frac{1}{2\mu}\left[ \frac{\partial^2}{\partial r^2} + \frac{1}{r^2 \sin \theta} \frac{\partial}{\partial \theta} \left( \sin \theta \frac{\partial}{\partial \theta} \right)  %
 - \frac{m^2}{r^2 \sin^2 \theta} \right] \chi_m + V(r,\theta)\chi_m = E_m \chi_m.
\end{align}
Multiple solutions of $E_m$ and $\chi_m(r,\theta)$ can be found by diagonalizing a discrete variable representation 
\cite{light_discrete-variable_2000}
using a basis of Hermite polynomials in the radial coordinate and Legendre polynomials in $\cos\theta$.
Finally, the ground-state tunnelling splitting is extracted from the difference between the lowest two energy levels within the $m=0$ manifold.

\end{document}